\documentclass[12pt]{article}
\usepackage{amsmath}
\usepackage{graphicx,psfrag,epsf}
\usepackage{enumerate}
\usepackage{apacite}
\usepackage{natbib}
\usepackage{url}
\usepackage{amsmath,latexsym,amssymb,amsthm}
\newcommand{\blind}{1}
\usepackage{sectsty}
\usepackage{layouts}
\usepackage{epigraph,xcolor}
\usepackage{graphicx}
\usepackage{pdfpages}

\usepackage{hyperref}

\hypersetup{colorlinks=true,citecolor=blue,backref,
	pdfpagemode=FullScreen,linktocpage=true}

\usepackage{setspace}
\begin{document}
\date{}

\if 1\blind
{
  \title{\bf The Statistical Face of a Region under Monsoon Rainfall in Eastern India}
 \author{Kaushik Jana$^1$, Debasis Sengupta$^2$, Subrata Kundu$^3$, \\Arindam Chakraborty$^4$ and Purnima Shaw$^5$ \\	$^1$ Imperial College London, UK\\$^2$Indian Statistical Institute, Kolkata, India\\    $^3$George Washington University, Washington, D.C., USA\\ $^4$Indian Institute of Science, Bangalore, India\\    $^5$Reserve Bank of India, Mumbai, India}
  \maketitle
} \fi

\bigskip

\begin{abstract}
A region under rainfall is a contiguous spatial area receiving positive precipitation at a particular time. The probabilistic behavior of such a region is an issue of interest in meteorological studies. A region under rainfall can be viewed as a shape object of a special kind, where scale and rotational invariance are not necessarily desirable attributes of a mathematical representation. For modeling variation in objects of this type, we propose an approximation of the boundary that can be represented as a real valued function, and arrive at further approximation through functional principal component analysis, after suitable adjustment for asymmetry and incompleteness in the data. The analysis of an open access  satellite data set on monsoon precipitation over Eastern Indian subcontinent leads to explanation of most of the variation in shapes of the regions under rainfall through a handful of interpretable functions that can be further approximated parametrically. The most important aspect of shape is found to be the size followed by contraction/elongation, mostly  along two pairs of orthogonal axes. The different modes of variation are remarkably stable across calendar years and across different thresholds for minimum size of the region.

\end{abstract}

\section{Introduction}\label{Introduction}
How large is the contiguous area under a particular spell of rainfall at a particular time? What is the typical shape of a raining cloud system? How does the shape vary from one spell to another?

While answers to these questions should be interesting by their own right, there could be other motivations for obtaining them as well. Rainfall systems with different sizes, shapes and orientations have received attention from meteorologists over the past few decades (see for example \cite{Hardenberg} and references cited there). In particular, precipitation areas associated with Mesoscale Convective Systems (MCS) are categorized according to their horizontal extent \citep{Austin_1972, Houze_2004}. These systems are major contributors to severe weather and precipitation in most parts of the earth \citep{Jun,Wertman}. In contrast to individual convective clouds that have scales from 1 to 10 km, MCS organize at 200 to 2000 km spatial scales \citep{Houze_1977, Jirak_2003}. Understanding of the distribution of shape, size and orientation of MCS may help us to improve Numerical Weather Prediction models, to study different covariate effects and to collate satellite data with terrestrial data such as radar measurements. Although there are several studies on the size of MCS, systematic accounts of their typical shape, size and orientation are not available. 

An opportunity to fill this gap arises from satellite data on rainfall rate, captured from 1997 to 2015 by the Tropical Rainfall Measuring Mission (TRMM), which focused on the region spanning from latitudes $50^\circ$N and $50^\circ$S. The data were collected through imaging over large swaths of area during the flight of a satellite. The images represent rainrate profile over the area covered at the time of recording, from which one can easily identify contiguous regions receiving positive rainfall at a particular instant.

The Eastern part of the Indian subcontinent is one of the wettest places of the world, and receives most of its precipitation during the South-Western monsoon season (also known as the monsoon season). The monsoon rainfall is extremely important for the agricultural sector of the economy in this region \citep{Kumar2004}. For this reason, prediction of precipitation pattern is also very important here. These circumstances make the monsoon rainfall of this region an ideal choice for the study mentioned above. 

The goal of this paper is to statistically model and analyze the spatial extent of a contiguous region under rainfall, by using a part of the above data that relates to the monsoon season in Eastern India. For this purpose, one has to assess the distribution of a suitably chosen representation of a region under rainfall. A rotationally invariant representation is not appropriate, since directionality is an important meteorological aspect of a rainfall system. Scale invariance is also undesirable, as there is no reason to presume that shapes of regions at different scales would have a similar pattern.

The problem of describing a spatial region falls in the domain of shape analysis. Many shape modelling techniques have been developed over the years with different advantages and drawbacks \citep{Rabiner_89, Loncaric_98}. Region based shape models generally involve representation through a binary matrix of pixels, which can be computationally cumbersome. The other approach is to describe shape objects in terms of their boundary contours, either through a finite number of landmarks or through a functional representation. Landmark based methods \citep{Mardia_98}, as well as functional representations adapted to landmarks \citep{Strait_2017}, generally require labelling of some features of the contours as landmarks, with domain-specific knowledge. A functional representation of the boundary contours (that does not depend on landmarks) is a more attractive option for automated processing.

In a functional representation of the boundary, a contour is typically described through a pair of continuous coordinate functions $x(t)$ and $y(t)$, where the parameter $t$ takes value over the periphery of the unit circle. These functions are not uniquely represent a contour, as an alternative pair of functions for the same contour can be obtained through a strictly increasing function of $t$ (i.e., a reparameterization). Uniqueness can be ensured through a specific parameterization, such as the arc-length parameterization \citep{Klassen_04}. However, the resulting representation does not result in a linear space, which would be ideal for probabilistic description. \cite{Srivastava_11} proposed a `square-root velocity function' (SRVF) representation together with a suitable metric for comparing curves. \cite{Kurtek_2012} provided a general framework for representing and analyzing size and orientation of curves together with shape, by building on the method provided by \cite{Srivastava_11}. If this approach is used for contours (closed curves), every contour would be represented by an equivalence class of vector-valued functions (i.e., SRVFs). These functions form an $L_2$ space, where the central contour and variations around it are typically described through the Karcher mean and the Karcher covariance function \citep{Kurtek_2012,Srivastava_2016}. It is known  \citep{Hall_2010} that such functional objects cannot have a density. Any notion of distribution is typically based on finite dimensional representations. However, an approximate representation of the SRVFs of the contours can be made through functional principal component analysis (FPCA) of the Karcher covariance function, followed by the use of the dominant principal components as a basis. We would see in Section~\ref{reconstruction} that this method requires hundreds of principal components for any meaningful representation of the contours of the regions under rainfall, which affects its performance.

In view of these difficulties of exact functional representation of the contours, one might look for an approximate representation. In this paper, we propose an envelope approximation of the boundary of a contiguous region under rainfall, which focuses on the directional aspect of its shape. The proposed representation is in terms of a single real valued function defined over the unit circle. It enables us to analyze these infinite dimensional objects by using standard statistical tools for functional data in a  Hilbert space \citep{Ramsay_05}, such as FPCA. A major difficulty is posed by censoring of the regions under rainfall by the field of view of the satellite. Based on preliminary findings of the analysis of the motivating data set, we propose an adjustment to the FPCA for incorporating censored regions. This adjusted FPCA leads us to a reasonably small set of  basis functions that explain most of the variation present in the TRMM rainfall data. We identify certain interpretable modes of variation present in the data. The analysis culminates in a parametric representation of the shape objects with only a few parameters that are readily interpretable. The values of the estimated parameters across the contours also reveal interesting patterns.

\section{Data and Pre-processing}
\subsection{TRMM Rainfall Data}\label{sec:trmm_data}

The TRMM data (TRMM~2B31, version 7), gathered by the US National Aeronautics and Space Administration (NASA) and the Japanese Aerospace Exploration Agency (JAXA), were derived by processing images recorded by the combined Precipitation Radar (PR) and TRMM Microwave Imager (TMI). The data can be freely downloaded from the Goddard Distributed Active Archive Center's website (http://trmm.gsfc.nasa.gov/). Since 2002, the horizontal resolution of the satellite was 5 km and swath width was 247 km. Each row of the data consists of the latitude and the longitude of a location representing the center of a 5 km by 5 km cell of a rectangular grid, average precipitation rate over that cell and the time of recording. The different rows correspond to different locations and/or time.

As mentioned in the Introduction, we consider rainfall data only for the monsoon season (June to September) during the years 2002 to 2012, and only a range of latitudes and longitudes in the Eastern part of the Indian subcontinent ($21^\circ$N-$30^\circ$N and $84^\circ$E-$90^\circ$E). Further, in order to focus on MCS, we consider contiguous regions of size 200 square kilometers (about eight pixels) or more, having positive rainfall. It may also be noted that directionality is an important feature in the study of regions under rainfall and the directionality of small regions cannot be studied meaningfully because of the limitation in resolution. On the other hand, overly large precipitation areas are not fully captured by the images, as noted in Section~\ref{Introduction}. In view of this limitation, regions with size larger than 13,500 square kilometers were also disregarded. In summary, the objects of study were limited to contiguous regions under rainfall having size 200 to 13,500 square kilometers, occurring during the monsoon months of the calendar years 2002 to 2012, over latitudes $21^\circ$N-$30^\circ$N and longitudes $84^\circ$E-$90^\circ$E. The total number of rainfall regions in the above time and spatial range is 15802 of which 11140 are complete (fully observed regions) and 4662 are incomplete (partially observed regions).

\subsubsection{An Example}
\begin{figure}
	\centering
	\includegraphics[height=6 in]{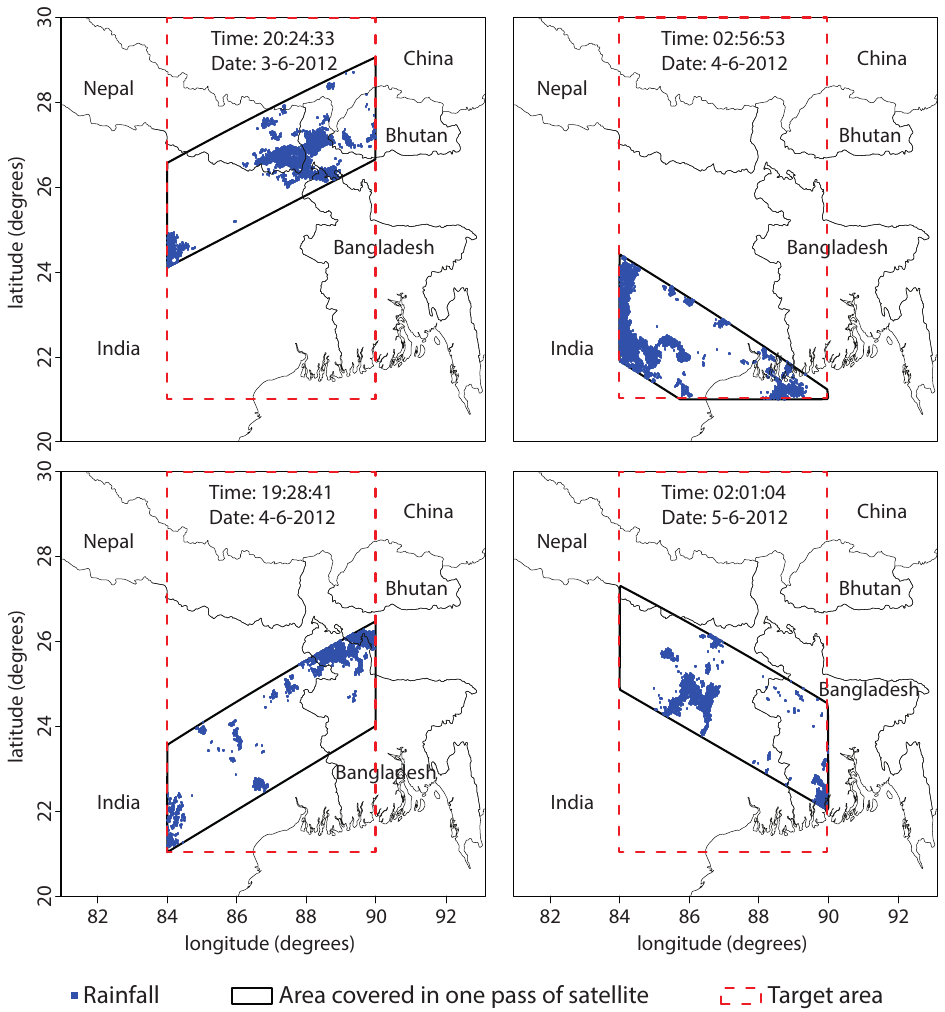}
	\caption[]{Target area, area covered and precipitation areas overlaid on the political map of the region, for four consecutive passes of TRMM satellite in June 2012.} \label{fig1}
\end{figure}

The four panels of Figure~\ref{fig1} show locations of rainfall (dark patches) as recorded by four consecutive passes of the TRMM satellite in the month of June 2012 over the target region (dashed boundaries), together with the areas covered in each pass (solid boundaries). The measured rate of rainfall at these locations are not shown in these figures. It is seen that these locations form spatial clusters.  Since each pass of the satellite covers the entire range of observations in a matter of seconds, the clusters obtained from a particular pass may be regarded as a single snapshot of the covered region at a particular instant.

\subsubsection{Regions under Rainfall and their Contours}\label{sec:region}
Even though the actual set of acquired data corresponds to locations lying on a grid, the position and spatial orientation of the grid depend on the varying flight path of the satellite during data acquisition. For image data acquired through a single pass of the satellite, a contiguous region under rainfall at a particular instant may be described as a set $S$ consisting of a cluster of grid cells $(i,j)$, such that
\begin{enumerate}
	\item for each $(i,j)\in S$, the rainfall rate is strictly positive;
	\item one can reach any element $(i,j)\in S$ from any other element $(i',j')\in S$ through a series of
	neighboring cells (including diagonal neighbors) which are also in $S$;
	\item for any $(i,j)\notin S$ that has a neighboring element in $S$, the observed rainfall rate is zero.
\end{enumerate}
The center of each grid cell is represented by its latitude and longitude. To find the boundary contour of a contiguous region under rainfall, we need to obtain the latitudes and the longitudes of the corners of these cells. In general, for a particular cell on the grid, there are four diagonal neighbors. The mid-points of the lines joining the center of the given cell and its diagonal neighbors are regarded as four vertices of the cell. When some neighbors are missing, a rectangular grid approximation is used locally. The non-self intersecting polygon obtained by joining the outer sides of the cluster of cells corresponding to a contiguous region under rainfall represents the boundary contour of that region.

There are more sophisticated techniques for boundary detection (see~\cite{Szeliski_2016}, \cite{Ghosal_2017} and \cite{Syring_2017}), which we do not use here. The methodology developed in this paper can be readily applied to contours estimated by these methods.

\subsubsection{Planar Projection}
The vertices of the above non-self intersecting polygons (referred to simply as polygons) are expressed in terms of latitude ($\beta$) and longitude ($\xi$), which describe location on the curved surface of the earth. These locations can be mapped onto a flat surface through any of the standard methods of projection \citep{Kells_1940}. We project all coordinates to the Euclidean space as follows. For any set $S$ as described in Section~\ref{sec:region}, we define its bounding region on the surface of the earth as $$B_S=[\beta_{min},\beta_{max}]\times[\xi_{min},\xi_{max}],$$
where
\begin{eqnarray*}
	&&\hskip-20pt \beta_{min}=\min\{\beta:\ (\beta,\xi)\in C_S\};\quad \beta_{max}=\max\{\beta:\ (\beta,\xi)\in C_S\};\\
	&&\hskip-20pt \xi_{min}=\min\{\xi:\ (\beta,\xi)\in C_S\};\quad \xi_{max}=\max\{\xi:\ (\beta,\xi)\in C_S\}.
\end{eqnarray*}
and $C_S$ is the set of latitude-longitude pairs $(\beta,\xi)$ of the boundary contour of $S$.

The center of the above bounding region is at $((\beta_{min}+\beta_{max})/2,(\xi_{min}+\xi_{max})/2)$. A projection of a general point within the bounding region, having latitude $\beta$ and longitude $\xi$ expressed in degrees, on a flat surface with origin (0,0) is given by the $x$ and $y$ coordinates (in kilometers) \citep{Kells_1940}:
\begin{eqnarray}
&&x=R\cos\left(\beta\frac{\pi}{180}\right)\left(\xi-\frac{\xi_{min}+\xi_{max}}{2}\right)\frac{\pi}{180}\nonumber,\\
&&y=R\left(\beta-\frac{\beta_{min}+\beta_{max}}{2}\right)\frac{\pi}{180},
\end{eqnarray}
where $R$ is the radius of the earth in kilometers.

Thus, the objects that would eventually be used for the requisite analysis are polygons on a planar surface, represented by the series of $x$ and $y$ coordinates of its vertices.

\subsection{Star-hull Representation}
Consider the problem of representing these polygons (contours), which can have irregular shapes. It should be noted that a rectangular region under rainfall with elongation along the North-South axis is different from another rectangular region with elongation along the East-West axis. The two regions may be outcomes of very differently behaving convective systems. Different regions under rainfall have to be distinguished in terms of their size as well. Thus, scale and orientation of the contours should not be standardized before they are compared. Therefore, in this special problem, the issue of alignment (registration) of the contours reduces to identification of a reference point for each contour. It could be the centroid of the enclosed region, the centroid of its convex hull, or any other uniquely defined point of physical importance. Without loss of generality we can align the reference point with the origin and denote it as $O$. For $0\leq\theta\leq 2\pi$, the set of intersections of the $\theta$-rays emanating from the reference point ($O$) with the contour contains all the information about that contour. The region bounded by the contour is called star-shaped with respect to $O$, when every $\theta$-ray has a unique point of intersection with the contour. Thus, a star-shaped contour can be represented simply through the set of radius vectors \citep{Kindratenko_2003} from the origin $O$ to the contour at different values of $\theta$. This simple representation is not possible for contours that are not star-shaped.

In order to permit a representation of all contours, we propose to use a star-shaped hull of any contour and model this hull. For a given contour and a reference point $O$, this hull would be defined as the locus of the furthest (from $O$) intersection of a $\theta$-ray with the contour for $\theta\in[0,2\pi]$. This locus would be a non-negative valued periodic function with period $2\pi$ which we call the `radial function'. One can use any convention for measuring $\theta$ (e.g., counter clockwise from the eastern direction so that $\theta=0$ lies to the east of the origin $O$) and follow it for all the contours. The star-hull would be the same for all such choices. Further, every star-hull would correspond to a unique radial function.

Rather than choosing the reference point heuristically, one might opt for an optimal choice. \cite{Arkin_98} proposed that one should use the hull with the minimum area, and termed it the star-shaped hull. They also found that the problem of identifying the star-shaped hull of a polygon with $n$ vertices, i.e., the search over the optimum reference point, takes $O(n^2)$ computational time. It is not clear though that this computation is a worthwhile exercise, as the optimal location of the reference point is not guaranteed to be unique or stable. Using a uniquely defined reference point, such as the centroid of the area or the centroid of the convex hull, might work better. We opt for the latter choice, as it produces a star-shaped-hull that is necessarily contained in the convex hull. We refer to the star-shaped hull corresponding to this heuristically chosen reference point as the star-hull of the original contour.

The star-hull approximation highlights the directional aspect of shape, which is very important in meteorological applications~(see Chapter~9 of \cite{Markowski_2010}). This simplification was also adopted by \cite{Micheas_2007}, who used farthest points on an angular grid as `landmarks' to represent a contiguous region under rainfall. The boundary estimation method of \cite{Ghosal_2017} is meant for star-shaped contours.

\begin{figure}
	\centering
	\includegraphics[width=6 in]{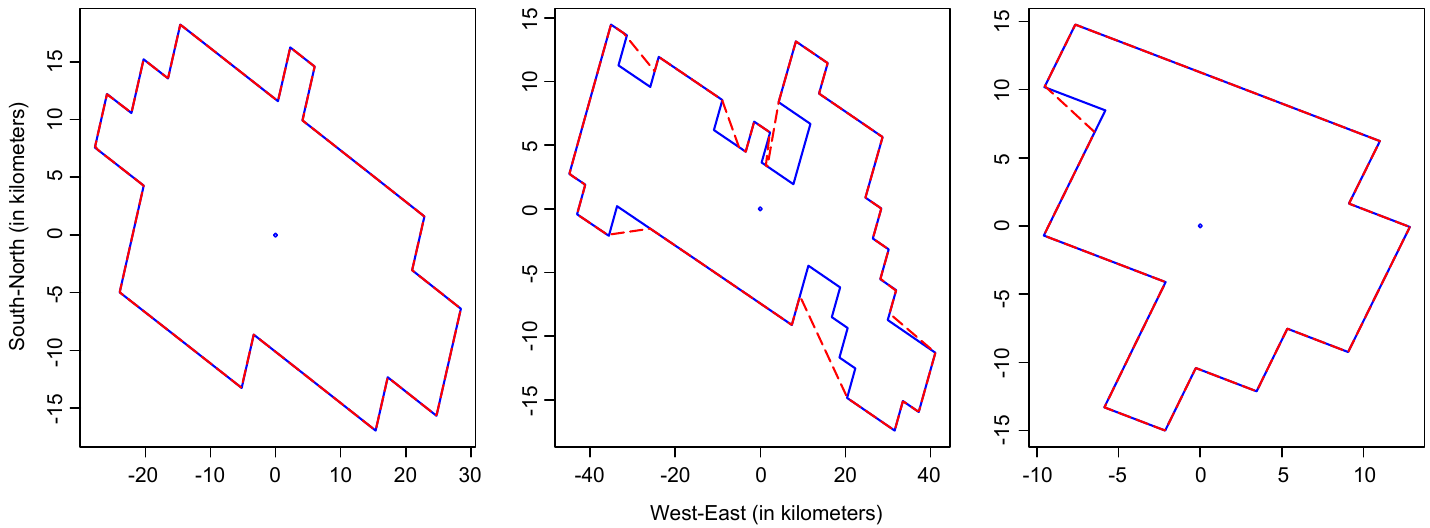}
	\caption[]{Examples of actual contours (solid line) and their star-hulls (dashed line), with the reference point (center of convex hull) marked as a dot.}\label{3.2}
\end{figure}

Figure~\ref{3.2} shows some examples of real contours from the TRMM data and the corresponding star-hulls, where the center of the convex hull is used as the reference point.

Next, we examine the accuracy of the star-hull approximation for the TRMM rainfall data. The computations necessary for this study are made on the basis of $\theta$-rays, over a uniform grid of 1000 values of $\theta$ over [0,2$\pi$]. (This grid size was chosen after verifying that grid sizes of 500 and 2000 produce nearly identical results.) The overall percentage error of the star-hull representation is measured by the area inside the star-hull not belonging to the region under rainfall, expressed as percentage of the area under rainfall. When this error is computed for all complete contours used in the present analysis (see Section~\ref{sec:trmm_data}), the three quartiles of the values happen to be 0.71\%, 2.60\%\ and 6.37\%. The total measure of angles $\theta$ over $[0, 2\pi]$, where the $\theta$-ray has multiple intersections with the actual contour (i.e., there is non-zero approximation error), has the three quartiles as $0.044\pi$, $0.136\pi$ and $0.284\pi$ (2.2\%, 6.8\%\ and 14.2\%\ of $2\pi$, respectively).

\begin{figure}[]
\centering
\includegraphics[width=5 in]{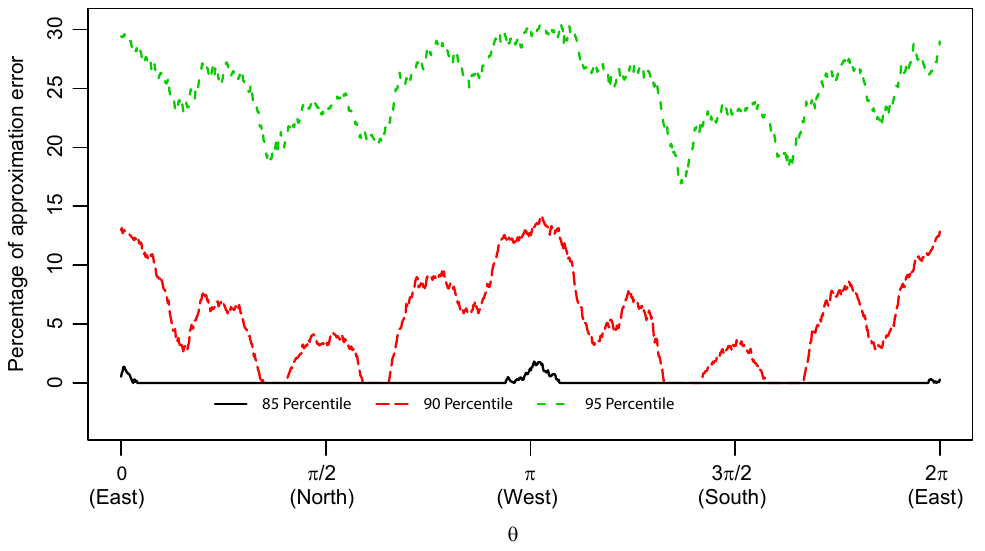}
\caption[]{Some percentiles of the star-hull approximation error in different directions for TRMM data, 2002-2012.}\label{fig2}
\end{figure}

Figure~\ref{fig2} gives an idea of the approximation errors in various directions, computed on the basis of the line joining the center of the star hull and a point on its periphery. For any given direction, this error is measured as the percentage of the line segment that is not contained in the precipitation area. All the three quartiles at all directions are zero. Three large percentile points of the error are plotted. The 90th percentile is on the average 5.5\%. These errors are much smaller than the overall approximation error in the finite dimensional representations presented in this paper (see Figure~\ref{fig12}).

Some directionality in the star-hull approximation error is also evident from Figure~\ref{fig2}, with largest error occurring along the East and West directions. This implies that most of the hollowness of the star-hull approximation happens in these directions. The other directions of larger error occur approximately at the interval of $\pi/4$. These findings may be used in future to embellish the model of the star-hull presented in this paper.

\subsection{Adjustment for Asymmetry}\label{sec:asymmetry}

\begin{figure}[]
\centering
\includegraphics[width=4.5 in]{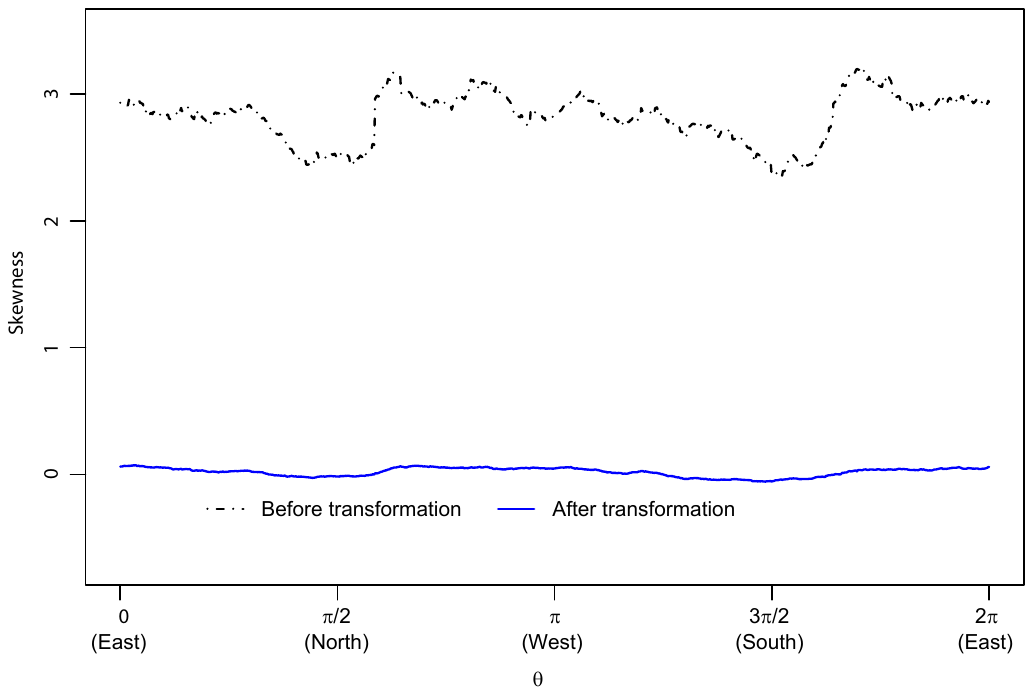} 	 
\caption[]{Direction specific skewness of star-hull radial distance for complete contours of analyzed TRMM data.}\label{fig4}
\end{figure}

A major challenge for the above analysis of the star-hull radial functions is the asymmetric distribution of the values of the function at any given point. The asymmetry arises from the fact that smaller contours are generally more abundant than larger ones in the above data. To examine the extent of this asymmetry, let us consider the set of relevant complete contours in the TRMM data (see Section~\ref{sec:trmm_data}). The set of values of the radial function for any given $\theta$, which indicates the size of the contours in that direction, have a very skewed distribution. This fact is brought out by Figure~\ref{fig4}, where the skewness is plotted against $\theta$ (dashed curve). Here the skewness at any direction $\theta$ is computed by using sample moments (i.e., by the ratio of the third central moment and the square root of the third power of the second central moment) of the values of the radial function at $\theta$ taking values over a uniform grid of size 1000 over the range $[0,2\pi]$.  For convenience of further analysis, we transform the data through a monotone function in such a way that the distribution of the transformed values of the radial functions (with all cases and angles pooled together) is standard normal. Specifically, the transformation is  $\Phi^{-1}\circ F_n$, where $\Phi$ is the standard normal distribution function, and $F_n$ is the empirical distribution function of the untransformed radial distances for all cases and all angles, chosen over a uniform grid of size 1000 over the range $[0,2\pi]$. The solid curve of Figure~\ref{fig4} shows the plot of the skewness of the transformed data against $\theta$. It transpires that this single transformation is able to bring the skewness at all directions close to zero. It is the transformed data that are used for the subsequent analysis.

\section{Non-parametric Modelling}
\subsection{Karhunen-Leove Expansion through FPCA}\label{FPCA}

We model the transformed radial function of each star-hull as the realization of a real valued stochastic process $X$ defined over the closed interval $[0,2\pi]$, with continuous and square-integrable sample paths satisfying $X(2\pi)=X(0)$. These sample paths form a Hilbert space, the inner product between $X$ and $Y$ being defined as $\int_0^{2\pi}X(\theta)Y(\theta)d\theta$. We assume that the stochastic process is second order stationary, with mean function $\mu:[0,2\pi]\rightarrow \mathbb{R}$ and covariance function $K:[0,2\pi]^2\rightarrow \mathbb{R}$. Assuming that $K$ is positive definite, it admits the spectral decomposition
\begin{eqnarray}\label{spec}
K(\theta,\theta')=\sum_{j=1}^\infty \lambda_j\varphi_j(\theta)\varphi_j(\theta'),\quad \theta,\theta'\in[0, 2\pi],
\end{eqnarray}
where $\lambda_1\geq \lambda_2\geq \cdots>0$ are the eigenvalues, with orthonormal eigenfunctions $\varphi_1,\varphi_2,\ldots$ respectively. The eigenvalues and the eigenfunctions may be identified through functional principal component analysis \citep{Ramsay_05, Ferraty_06} of the covariance function. The spectral decomposition leads us to the Karhunen-Loeve expansion of the process $X$ (see \cite{Bosq}) as
\begin{equation}\label{kl}
X(\theta)=\mu(\theta)+\sum_{j=1}^{\infty} X_{[j]}\varphi_j(\theta),\quad \theta\in[0, 2\pi],
\end{equation}
where $X_{[j]}=\int_0^{2\pi}(X(\theta)-\mu(\theta))\varphi_j (\theta)d\theta$ is the $j$th principal component of $X$.

Given a sample of $n$ radial functions $\{X_1(\theta),\ldots,X_n(\theta);\theta\in [0,2\pi]\}$, estimates of $\mu$ and $K$ may be obtained as
\begin{align}\label{sample_mean_cov}
\begin{split}
\hat{\mu}(\theta)&=\frac{1}{n}\sum_{i=1}^n X_{i}(\theta)\quad and \\
\hat K(\theta,\theta')&=\frac{1}{n}\sum_{i=1}^n\left(X_i(\theta)-\hat{\mu}(\theta)\right)
\left(X_i(\theta')-\hat{\mu}(\theta')\right),\;\; \theta,\theta'\in[0, 2\pi].
\end{split}
\end{align}
The eigen elements of $\hat K$ provide estimators of $\lambda_j$ and $\varphi_j$ as $\hat\lambda_j$ and $\hat\varphi_j$, respectively, for $j=1,2,\ldots$, which lead to a sample version of the Karhunen-Leove expansion. The finite number of principal components to be retained is determined from the usual consideration of explaining most of the variability in the data.

\subsection{Handling of Censored Data}\label{sec:censored}
During a particular pass of the satellite, a part of a  contiguous region under rainfall may fall outside the swath covering it. One does not have adequate information about the shape or size of such a region. This is a form of spatial censoring. The relevant part of the TRMM data (see Section~\ref{sec:trmm_data}) contains nearly 30\%\ censored regions. Treating them as complete ones would lead to biased estimates. On the other hand, ignoring them would lead to inefficient estimates.

We would observe in Section~\ref{sec:FPCA_sample} that the largest component of the variation present in the data is the variation in size of the region under rainfall. We use this crucial fact to handle censored data.

Consider the area within each of the star-hulls of complete and censored regions under rainfall as realizations of a positive valued random variable. Some of these observations (where contours are fully observed) are complete observations, and some others (where contours are partially observed) are censored. If the censored contours had been fully observed, the area enclosed by them would have been larger. Therefore, the sizes of the contours may be regarded as right-censored data. The size distribution can be estimated by $1-\hat S(s)$, where $\hat S(s)$ is the Kaplan-Meier estimator of the survival function, 
\begin{equation}
\hat S(s)= \left\{
\begin{array}{ll}
1 & \mbox{if}\; s < s^*_1,\\
\frac{1}{n}\prod_{i=1}^m\left(1-\frac{\sum_{j=1}^n \delta_jI(s_j=s^*_i)}{\sum_{j=1}^n I(s_j\geq s^*_i)}\right)^{I(s^*_i\leq s)} & \mbox{if}\; s \geq s^*_1,
\end{array} \right.
\end{equation}
where $s_1,\ldots, s_n$ are the contour sizes, $s^*_1,\ldots, s^*_m$ are ordered distinct complete contour sizes, $I(\cdot)$ is the binary indicator of the event appearing in its argument and $\delta_i=I(i^{th}$ contour is complete), $i=1,\ldots,n$. The corresponding estimator of the distribution function, $1-\hat S(s)$, has (possibly unequal) positive probabilities allocated to the complete cases, and zero probability mass to the censored cases. Let $\hat S(s-)$ be the left limit of $\hat S$ at $s$ and $p_i=\hat S (s_i-)-\hat S (s_i)$, i.e., the probability mass allocated by the Kaplan-Meier estimator to the $i$-th contour, $i=1,\ldots,n$. The mean and the covariance functions are
then estimated as
\begin{align}\label{mean_cov_censored}
\begin{split}
\hat{\mu}(\theta)&=\sum_{i=1}^{n}p_i X_i(\theta),\;\text{and} \\
{\hat K(\theta,\theta')}&=\sum_{i=1}^{n} p_i \left(X_i(\theta)-\hat{\mu}(\theta)\right)\left(X_i(\theta')-\hat{\mu}(\theta')\right), \theta,\theta'\in [0,2\pi].
\end{split}
\end{align}
The eigenvalues and eigenfunctions $(\lambda_j,\varphi_j)$, $j=1,2,\ldots,$ are then estimated by replacing
\eqref{sample_mean_cov} with \eqref{mean_cov_censored}.

It may be possible to adapt the above method to utilize the partial shape information of a censored contour. While the pool of complete contours `matching' the incomplete one may be rather limited (in the absence of size and rotational invariance), one can use proximity-based weights for candidate contours. Whether the additional sophistication would bring any worthwhile benefit remains to be seen. One can also use parametric and/or Bayesian imputation (see \cite{Gu_2014}). Appropriate parametric models for the present application, once identified, may be used in future to look for better ways of handling censored data.

\subsection{FPCA for TRMM Rainfall  Data}\label{sec:FPCA_sample}

For the TRMM contours data described in Section~\ref{sec:trmm_data}, we use their star-hull radial functions for FPCA, after transforming the radial distances as mentioned in Section~\ref{sec:asymmetry}.

\begin{figure}[]
	\centering
	\includegraphics[width=5 in] {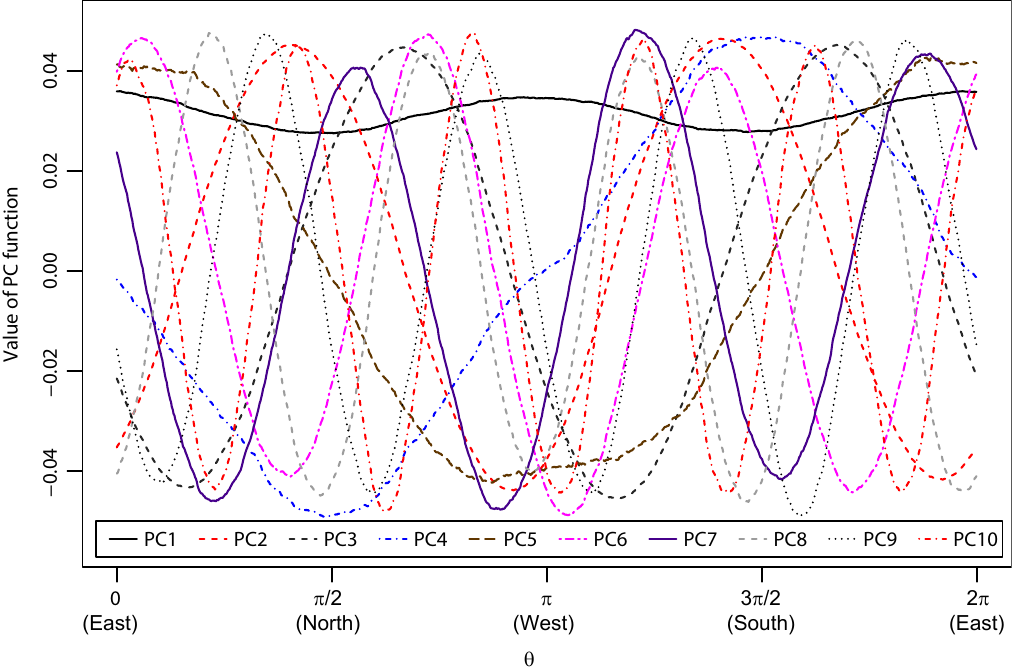}
	\caption[]{Plots of the first ten principal component basis functions, based on complete contours over the years 2002-2012.}\label{pc14}
\end{figure}

\begin{table}
\caption{Percentage of variance explained by the first $j$ principal components, for $j=1,\ldots,12$, in the star-hull contours of regions under rainfall, based on complete contours over the years 2002-2011.}\label{table_3.1}
\vspace{0.5cm}
\centering
\begin{scriptsize}
			\begin{tabular}{ccccccccccccc}\hline
				$j$    & 1 & 2&3&4&5&6&7&8&9&10&11&12\\\hline
				$\lambda_j$& 35.85& 52.76& 68.77& 75.57& 80.93& 84.04& 87.01& 89.00& 90.95& 92.03&93.08& 93.78\\ \hline
			\end{tabular}
	\end{scriptsize}
\end{table}
For preliminary analysis, we consider only the complete contours. The cumulative sums of the estimated eigenvalues ($\hat \lambda_j$) corresponding to the largest few principal components, expressed as percentage of the sum of all eigenvalues (i.e., the total variance) and computed from the monsoon data over the years 2002-2012, are reported in Table~\ref{table_3.1}. The eigenvalues decrease fairly quickly, with the first ten principal components accounting for 92 per cent of the total variation. In Figure~\ref{pc14}, we plot the first ten estimated principal component basis functions. Observe that the first principal component function (solid curve) is almost a constant, with weak extrema along East, West, North and South. It can be said that the first principal component, which alone accounts for about thirty five percent of the total variation, captures mostly the variation in size of the regions. The other components, which are orthogonal to the first component, capture different aspects of the variation in shape of the contours.

\begin{figure}[]
	\centering
	\includegraphics[height=2.25 in] {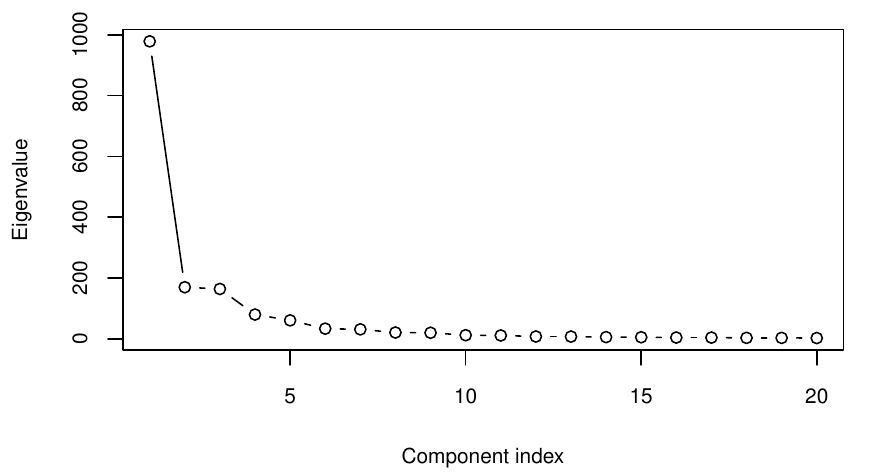}
	\caption[]{Scree plot of the eigenvalues obtained from the adjusted FPCA of contours of regions under rainfall over the years 2002-2012.}\label{scree_2011}
\end{figure}

We now turn to FPCA with adjustment for censoring. In this analysis, the estimates \eqref{sample_mean_cov} of the mean and the covariance functions are replaced by the estimates given in \eqref{mean_cov_censored}. We  present in Figure~\ref{scree_2011} the eigenvalues ($\hat \lambda_1,\ldots,\hat \lambda_{20}$) obtained by using the adjusted FPCA. The eigenvalues are seen to decrease rapidly. It turns out that the cross-validation score based on the integrated mean square of the leave-one-curve-out prediction error \citep{Rice_1991, Yao_2005}, pooled across the years is minimized when the first twelve eigenfunctions are used for approximation. The twelve corresponding principal components account for 95 per cent of the total variation.

\begin{figure}[t]
	\centering
	\includegraphics[width=4 in]{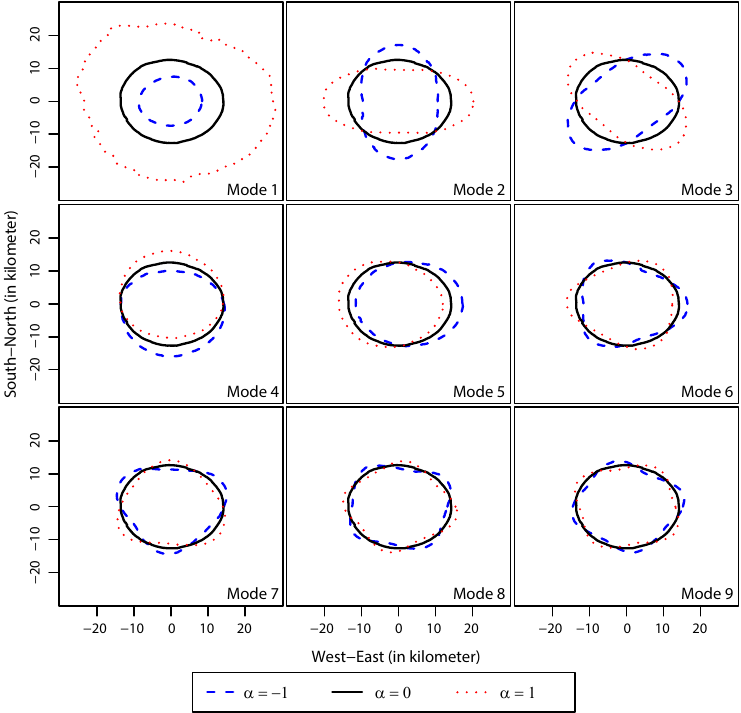}
	\caption[]{Plots of the first nine modes of variations of the regions under rainfall over the years 2002-2012.}\label{mv}					
\end{figure}

For visualization of the principal component functions, we use a specially designed plot. Consider the function
\begin{equation}\label{mv_definition}
\eta_{j,\alpha}(\theta) = \hat\mu(\theta) + \alpha \sqrt{\hat\lambda_j} \hat\varphi_j(\theta),\;\; \theta\in [0,2\pi],
\end{equation}
where $\alpha$ is a fixed number chosen from the interval $[-2,2]$. This function is generally referred to as the $j$th mode of variation (MV) \citep{JM}. The function $g\left(\eta_{j,\alpha}(\theta)\right)$, where $g$ is the re-transformation function $F_n^{-1}\circ\Phi$ (corresponding to the transformation function described in Section~\ref{sec:asymmetry}) and $F_n^{-1}(u)=\inf_{x}\{F_n(x)\ge u\}$, represents the $j$th MV in the original (re-transformed) scale. The plot of $g\left(\eta_{j,\alpha}(\theta)\right)$ against $\theta$ in polar coordinates would show a typical departure of the $j$th MV from the mean function. In order to visualize the $j$th MV, we overlay this graph for $\alpha=-1,0$ and 1. The overlaid graphs of the first nine MV are displayed in Figure~\ref{mv}.

In each of the nine panels of Figure~\ref{mv}, the case $\alpha=0$ corresponds to $\hat\mu(\theta)$ in the re-transformed scale.  This central region happens to be  approximately circular with a radius of about 10 km.

The first mode of variation (panel marked as Mode 1), represents the variation in size of the regions under rainfall. One standard deviation of departure from the median function in the inward direction leads to about 50\% reduction in size, while the same amount of departure in the outward direction leads to about 250\% increase in size.

The second MV (panel marked as Mode 2) represents contrasting departures from the median function in the North-South and the East-West axes. When $\alpha=1$, we have about 40\%\ elongation in East-West axis and about 20\%\ contraction in North-South axis. When $\alpha=-1$, there is reversal in the directions of contraction and elongation. Therefore, the second MV (the second most important aspect of the total variation) relates to whether the region under rainfall is elongated in North-South direction or in East-West direction, and how sharp that elongation is. This pattern of departure from the median function is less pronounced than the first mode of variation, shown in the top left panel.

The third MV (panel marked as Mode 3) is similar to the second MV, except that the axes of contrasting elongations and contractions are rotated by about $45^\circ$. The amount of departure from the median function is comparable to the typical departure observed in the case of the second MV.

The fourth MV (panel marked as Mode 4) shows departures from the median function that are confined mostly to  North and South directions. When $\alpha=1$, there is about 25\% elongation in North and about 15\% contraction in South. The pattern reverses when $\alpha=-1$. Thus, the fourth MV indicates whether the region under rainfall is pointed towards North or South. Overall, this feature is rather subdued in comparison to the cases of the first three MVs.

The fifth MV (panel marked as Mode 5) is similar to the fourth MV, except that the contrasting elongations and contractions are in East and West directions. The amount of departure from the median function is also comparable to the departure observed in the case of the fourth MV.

The sixth, seventh, eighth and ninth MVs, shown in the bottom panels, have similar interpretations.

\begin{figure}[]
\centering
\includegraphics[width=5 in] {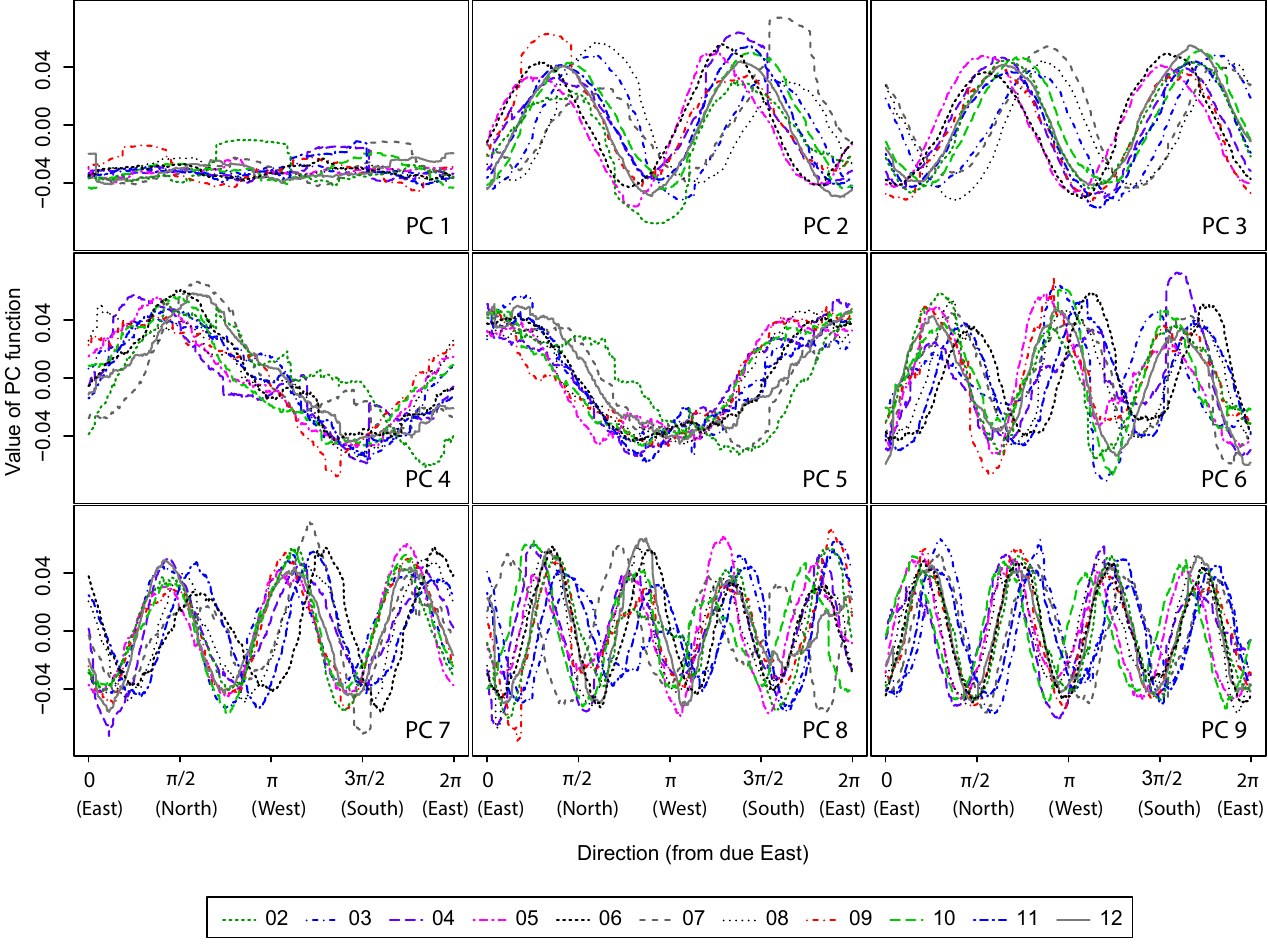}
\caption[]{Plot of the first nine PC functions corresponding to the years 2002-2012 using FPCA based on data of different years. Each panel corresponds to a single PC function, while each curve in a single plot represents a particular year (2002 to 2012).}\label{yr}
\end{figure}

In Figure \ref{yr}, we plot the principal component (PC) functions corresponding to the years 2002-2012 estimated by using proposed version of FPCA based on data of different years. Each panel corresponds to a single PC function, while each curve in a single plot represents a particular year (2002 to 2012). It is found that (i) the first PC function is approximately a constant, (ii) the fourth and the fifth PC functions constitute an orthogonal pair of approximately sinusoidal functions with period $2\pi$ (fundamental or lowest frequency), (iii) the second and the third PC functions constitute an orthogonal pair of approximately sinusoidal functions with period $\pi$ (first harmonic), (iv) the sixth and the seventh PC functions constitute an orthogonal pair of approximately sinusoidal functions with period $2\pi/3$ (second harmonic), (v) the eighth and the ninth PC functions constitute an orthogonal pair of approximately sinusoidal functions with period $2\pi/4$ (third harmonic).

This pattern is remarkably stable across the years. There is occasional change in phase from one year to another, but the shapes of the functions are consistent across the years. The variances of principal components associated with these functions, reported in Table~\ref{yr_value}, also show similar proportions of the total variation being explained by the different PC functions. The pairs of sinusoids of a particular frequency/period have about the same value of the associated variance of principal component.
\begin{table}
	\vspace{0.4cm}
	\caption{Percentage of variance explained by each of the first $j$ principal components, for $j=1,\ldots,12$. Each row represents eigenvalues corresponding to a particular year (2002 to 2012).}\label{yr_value}
	\vspace{0.5cm}
	\centering
	\begin{scriptsize}
		\begin{tabular}{ccccccccccccc}\hline
			&\multicolumn{12}{c}{Eigenvalue corresponding to PC number (in decreasing order)}\\
			\cline{2-13}
			Year    & 1 & 2&3&4&5&6&7&8&9&10&11&12\\\hline
			$2002$ & 50.18 & 17.40& 9.89& 5.96& 3.97& 2.39& 1.97& 1.22& 1.13& 0.79& 0.64& 0.46\\
			$2003$ &57.47& 10.90& 9.40& 4.81& 4.18& 2.99& 2.03& 1.32& 1.21& 0.84& 0.74& 0.49\\
			$2004$ &49.29& 12.72& 11.20& 8.22& 4.11& 2.88& 2.12& 1.48& 1.36& 0.93& 0.79& 0.54\\
			$2005$ &58.12& 12.22& 9.78& 4.34& 3.17& 1.98& 1.94& 1.37& 1.20& 0.75& 0.64& 0.4\\
			$2006$ &54.37& 11.77& 11.14& 6.04& 3.66& 2.25& 1.85& 1.41& 1.27& 0.74& 0.66& 0.61\\
			$2007$ &53.00& 15.68&  9.74& 5.11& 3.79& 2.49& 1.72& 1.52& 1.13& 0.77& 0.70& 0.50\\
			$2008$ &50.36& 12.59&  11.48& 6.48& 4.84& 2.38& 2.05& 1.56& 1.44& 0.82& 0.78& 0.60\\
			$2009$ &57.93& 11.40&  10.38& 4.60& 3.61& 2.35& 1.70& 1.24& 1.14& 0.75& 0.69& 0.47\\
			$2010$ &55.41& 12.53& 10.47& 4.29& 4.00& 2.38& 2.15& 1.51& 1.21& 0.73& 0.62& 0.56\\
			$2011$ &61.17& 10.20&  9.25& 5.01& 3.36& 1.83& 1.59& 1.11& 1.05& 0.65& 0.58& 0.47\\
			$2012$ &58.94& 10.86&  9.67& 5.16& 3.47& 2.06& 1.69& 1.22& 1.14& 0.80& 0.69& 0.42\\\hline
		\end{tabular}
	\end{scriptsize}
\end{table}

As mentioned in Section~\ref{sec:trmm_data}, the threshold for minimum size of a region under rainfall, used in the above analysis is 200 square kilometers. The modes of variation are found to be stable when this threshold is chosen anywhere between 50 and 1000 square kilometers.

\section{Parametric Modeling}\label{sec:parametric}

The PC functions described in the preceding section have close resemblance with sinusoids. The natural set of basis functions for this data appears to consist of sinusoids with integer frequency (i.e., Fourier basis functions). This coincidence is remarkable. In the present context, a sinusoid with integer frequency represents a finite number of equally spaced directions of elongation, interspersed with an equal number of directions of contraction. These basic shapes are attractive because of their simplicity and clear directionality. If contours are represented in terms of these shapes, and variation in shape can be decomposed into variation in these constituent shapes, that would be a welcome simplification.

With the aim of a Fourier series approximation of the contours, we consider the following functional linear model \citep{Ferraty_06} for the $k$th contour (after transformation as in Section~\ref{sec:asymmetry}),  $k=1,\ldots,n$,
\begin{eqnarray}\label{fourier1}
X_k(\theta)=\sum_{i=0}^{d_k} A_{ki}\cos(i\theta)+ \sum_{i=1}^{d_k} B_{ki}\sin(i\theta)+e_k(\theta),\quad \theta\in[0, 2\pi],
\end{eqnarray}
where $d_k$ is the order of the Fourier series approximation, $A_{ki}$, $B_{ki}$ are the coefficients of the sinusoids with frequency $i$ and $e_{k}(\theta)$ is a zero mean Gaussian stochastic error for the $k$th contour. A more compact description of the model is
\begin{eqnarray}\label{fourier2}
X_k(\theta)=\sum_{i=0}^{d_k} C_{ki}\cos(i\theta-i\phi_{ki})+e_{k}(\theta),\quad \theta\in[0,2\pi],
\end{eqnarray}
where $(C_{ki},i\phi_{ki})$ is the polar representation of the point with Cartesian coordinates $(A_{ki},B_{ki})$ for $i\ge1$ and $C_{k0}=A_{k0}$.

For the $k$th contour, the Fourier coefficients in \eqref{fourier1}, ${\bf A}_k=(A_{k0},\ldots,A_{kd_k})$ and ${\bf B}_k=(B_{k1},\ldots,B_{kd_k})$ were estimated by the least squares method and the values of $d_k$ were selected by using the risk based data driven technique (see \cite{Hart}, page 167). Both the mode and the median of $d_1,\ldots,d_n$ happened to be 6 based on the contours over the years 2002-2012. Therefore, we chose $d_k=6$ for all $k$. This selection corresponds to thirteen sinusoidal terms in the right hand side of \eqref{fourier1}, which is comparable to the twelve principal component functions that had been deemed adequate in Section~\ref{sec:FPCA_sample}.

Let $(\hat{\bf A}_k$, $\hat{\bf B}_k)$ be the least squares estimators of $({\bf A}_k,{\bf B}_k)$, described above, for $k=1,\ldots,n$. For each $k$, the parametric estimate of the $k$th contour is given by 
\begin{equation}\label{approx_para}
X_k(\theta;\hat{\bf A}_k,\hat{\bf B}_k)=\sum_{i=0}^{d} \hat A_{ki}\cos(i\theta)+ \sum_{i=1}^{d} \hat B_{ki}\sin(i\theta)=\sum_{i=0}^{d} \hat C_{ki}\cos(i(\theta-\hat\phi_{ki})), \quad \theta\in[0, 2\pi],
\end{equation}
where $d$ is the chosen order of the Fourier approximation, $\hat C_{k0}=\hat A_{i0}$, and $(\hat C_{ki},i\hat\phi_{ki})$ is the polar representation of the point with cartesian coordinates $(\hat A_{ki},\hat B_{ki})$ for $i\ge1$.
The equivalent order nonparametric estimate of the star-hull of the $k$th contour from the FPCA model \eqref{kl},
\begin{equation}\label{approx_nonpara}
\hat X_k (\theta)=\hat\mu(\theta)+\sum_{j=1}^{2d+1}\left(\int_0^{2\pi}(X_k(\zeta)-\hat\mu(\zeta))\hat\varphi_j(\zeta)d\zeta \right)\hat\varphi_j(\theta), \quad \theta\in[0, 2\pi].
\end{equation}

It should be recalled that these representations are in a transformed domain. The Fourier and FPCA approximations for the $k$th region in the original domain are the regions bounded by the contours  \begin{eqnarray}
&&g(X_k(\theta;\hat{\bf A}_k,\hat{\bf B}_k)),\quad\theta\in[0, 2\pi],\label{fourier3}\\
&&g(\hat X_k(\theta)),\quad\theta\in[0, 2\pi],\label{g3}
\end{eqnarray}
where $X_k(\theta;\hat{\bf A}_k,\hat{\bf B}_k)$ and $\hat X_k(\theta)$ are given by \eqref{approx_para} and \eqref{approx_nonpara}, respectively, and $g$ is the re-transformation function defined in Section~\ref{sec:FPCA_sample}.

\subsection{Error in Finite Dimensional Representation}
\begin{figure}[]
	\centering
	\includegraphics[width=5 in]{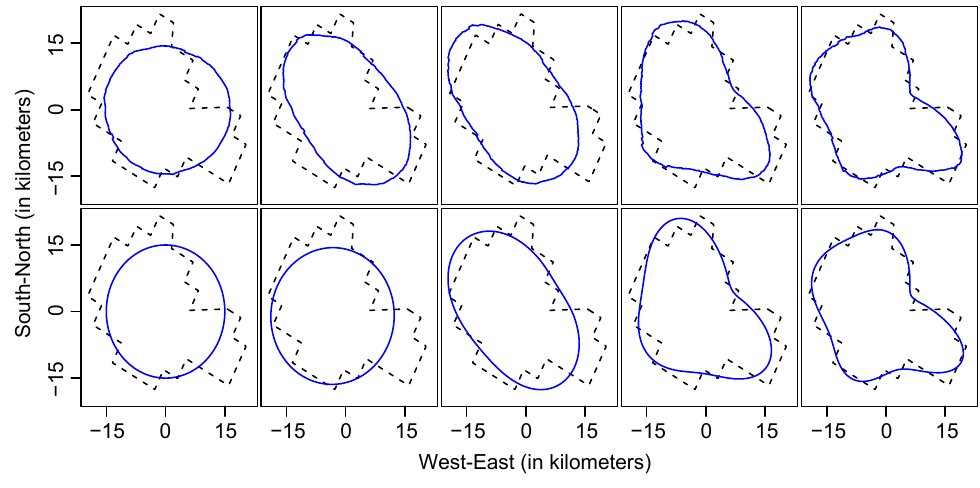}
	\caption[]{An example of reconstructed contour using cumulative totals of the first few PC (top row) and Fourier (bottom row) basis functions. The reconstructed contour is shown in solid lines, while the star-hull contour is shown in dashed lines.}\label{recons_1}
\end{figure}

As an illustration of the nature of approximation of a typical contour of regions under rainfall, we show in Figure~\ref{recons_1} the results of successively improved approximations. The top panel shows the contours of the nonparametric (FPCA) approximation $g(\hat{X}_k(\theta))$ defined in \eqref{g3} with $d=0,1,2,3,4$. 
The bottom panel shows the contours of the parametric (Fourier) approximation $g(X_k(\theta;\hat{\bf A}_k,\hat{\bf B}_k))$ defined in \eqref{fourier3}. The leftmost plot in each panel represents only one summand and the other plots represent successively higher number of terms in the summation. Inclusion of more and more summands evidently leads to better approximation. The Fourier series approximation is quite similar to the nonparametric one at almost every stage.

\begin{figure}[]
\centering
\includegraphics[width=6in] {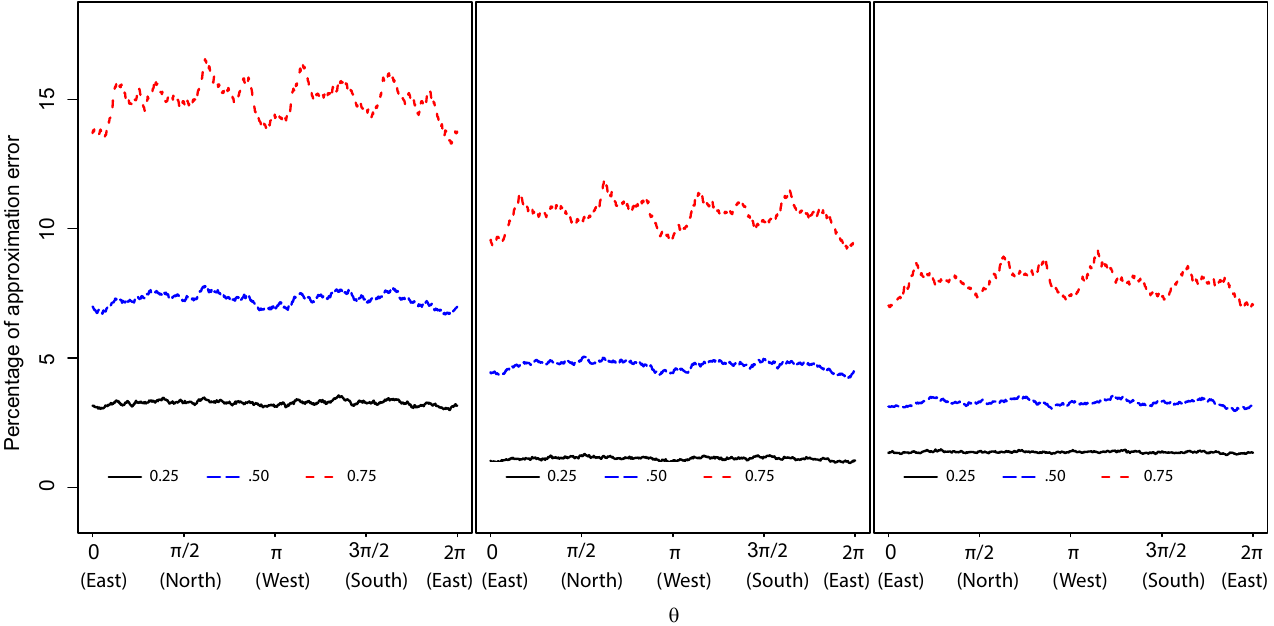}
\caption[]{Quartiles of normalized error of approximation (through \eqref{fourier3}) in different directions for regions under monsoon rainfall during 2002-2012. The left, middle and right panels corresponds to \eqref{fourier3} with $d=6,9$ and 12, respectively.}\label{fig12}
\end{figure}

Let us now examine the Fourier representation~\eqref{fourier1} with different values of $d$ in terms of the error in approximating the actual contour, whose star-hull had so far been approximated.  We focus on the direction specific approximation error for the finite Fourier series representation at each $\theta\in[0,2\pi]$, and define two sets:
\begin{align}\label{error1}
\begin{split}
	\delta_\theta&=\{x: (x,\theta) \mbox{ is contained in the actual region}\},\\
	\eta_\theta&=\{x: (x,\theta) \mbox{ is contained in the Fourier approximated region}\}.
	\end{split}
\end{align}
Note that $\eta_\theta$ is an interval, while $\delta_\theta$ can be either a single interval or a union of intervals. The symmetric difference between $\delta_\theta$ and $\eta_\theta$ (the set of points that lie either in $\delta_\theta$ or in $\eta_\theta$ but not in both; denoted here by $\delta_\theta\Delta\eta_\theta$) would in general be a unions of intervals. For a fixed $\theta\in [0,2\pi]$, let $l(\delta_\theta)$ and $l(\delta_\theta\Delta\eta_\theta)$ be the total lengths of the unions of intervals represented by $\delta_\theta$ and $\delta_\theta\Delta\eta_\theta$, respectively. We define the percentage of Fourier series approximation error in the direction $\theta$ as $p_\theta=\frac{l(\delta_\theta\Delta\eta_\theta)}{l(\delta_{\theta})}\times100$, and plot it against $\theta$, for $d=6$, 9 and 12 and for 1000 uniformly spaced values of $\theta$ in Figure~\ref{fig12}. The figure shows how the error reduces with greater number of Fourier series terms included in the representation. There is no dominant directionality in the error.

\subsection{Distribution of the Approximated Regions under Rainfall}

\begin{figure}[]
	\centering
	\includegraphics[width=5.5 in] {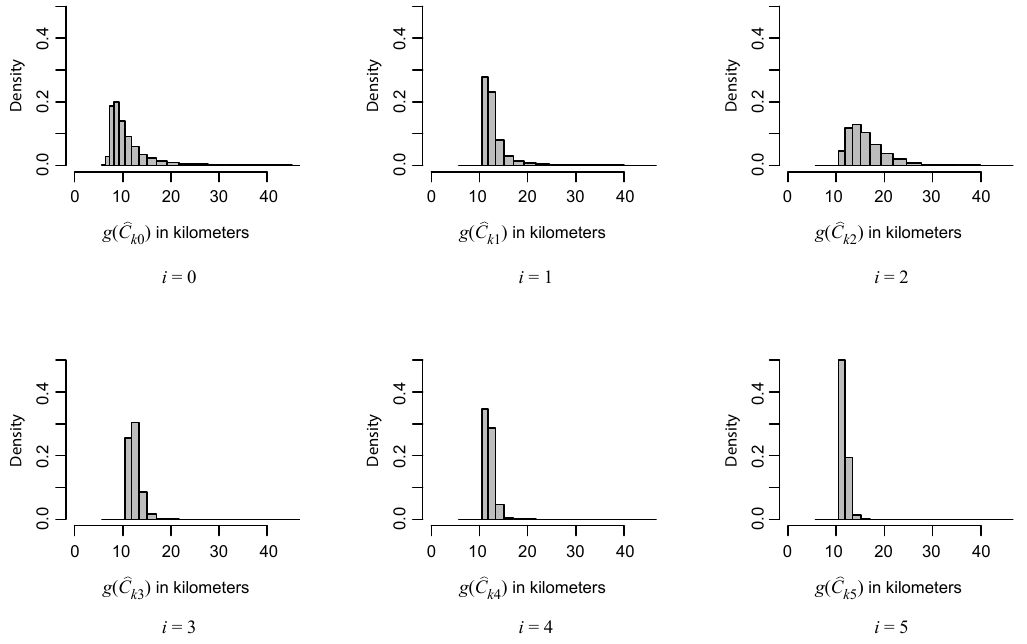}
	\caption[]{Histograms of re-transformed amplitudes of different sinusoids of \eqref{approx_para}. The top left, middle and right panels corresponds to $i=0$, 1, and 2 while the bottom left, middle and right panels indicate $i=3$, 4, and 5, respectively.}\label{amplitude1}
\end{figure}

Once the parametric representation of \eqref{fourier2} is chosen, the star-hull contour of every region under rainfall is represented by a set of amplitude and phase parameters. Therefore, the distribution of these shape objects is approximately described by the $(2d+1)$-variate distribution of these amplitudes and phases of the sinusoids. In order to appreciate different aspects of this distribution, we consider the marginal distributions of the amplitudes of the different sinusoids, represented in the original (re-transformed) scale. Specifically, for $i=0,1,\ldots,d$, we represent the contribution of the $i$-th summand in \eqref{approx_para} by $g\left(\hat C_{ki}\right)$. In Figure~\ref{amplitude1}, we have plotted the histograms of the first six re-transformed amplitudes computed from all complete contours used in Section~\ref{FPCA}. It is not surprising that the distributions of the re-transformed amplitudes are positively skewed. The histogram of the constant term $g(\hat C_{k0})$ (panel $i=0$ of the figure) has a smaller mode. This is because of the fact that $\hat C_{k1},\ldots,\hat C_{k5}$ are constrained to be positive, while $\hat C_{k0}$ is not. About half of all the contours have $\hat C_{k0}<0$, and the re-transformed values of these are too small to be attainable by the re-transformed values of the other amplitudes ($\hat C_{k1},\ldots,\hat C_{k5}$). The histograms of $g(\hat C_{k1})$ and $g(\hat C_{k2})$ (($i=1$) and ($i=2$) of Figure~\ref{amplitude1}) have wider support than the other histograms in this figure, indicating that the corresponding sinusoids in \eqref{approx_para} have most diversity in amplitude.

The histogram $g(\hat C_{k2})$ (panel $i=2$ of Figure~\ref{amplitude1}) has the widest support, apart from $g(\hat C_{k0})$. The associated sinusoid approximately corresponds to the second and the third modes of variation identified in Section~\ref{sec:FPCA_sample}. It represents a pattern of elongation/contraction of the mean circular shape that captures the main axiality of the region under rainfall. To identify the modal axis, we consider the bivariate distributions of $(\hat A_{k2},\hat B_{k2})$. Each contour contributes a pair $(\hat A_{k2},\hat B_{k2})$. We estimate the mode of this distribution by using the kernel density based method of \cite{Abraham_2003} with the Gaussian kernel and optimal bandwidth as suggested by \cite{Scott_2008}. Let us denote $(\hat A^*_2,\hat B^*_2)$ as the estimated bivariate mode, expressed in polar coordinates as $(\hat C^*_2,2\hat \phi^*_2)$, where $\hat C^*_2$ is the modal amplitude and $\hat \phi^*_2$ is the modal phase or direction. Then the modal version of the second sinusoid in \eqref{fourier1} is $\hat C^*_2\cos(2(\theta- \hat \phi^*_2))$. The corresponding re-transformed contour is the graph of $g(\hat C^*_2\cos(2(\theta- \hat \phi^*_2)))$ against $\theta$ in polar coordinates. This contour represents the modal axiality of the region under rainfall, as captured by the second sinusoid (with period $\pi$) in \eqref{fourier1}. The length of diameter of this contour ($2\hat C_2^*$) describes the strength of the modal axiality, while the pair of angles $\hat\phi_2^*$ and $\hat\phi_2^*+\pi$ define its direction.

\begin{figure}[]
\centering
\includegraphics[width=3.5 in] {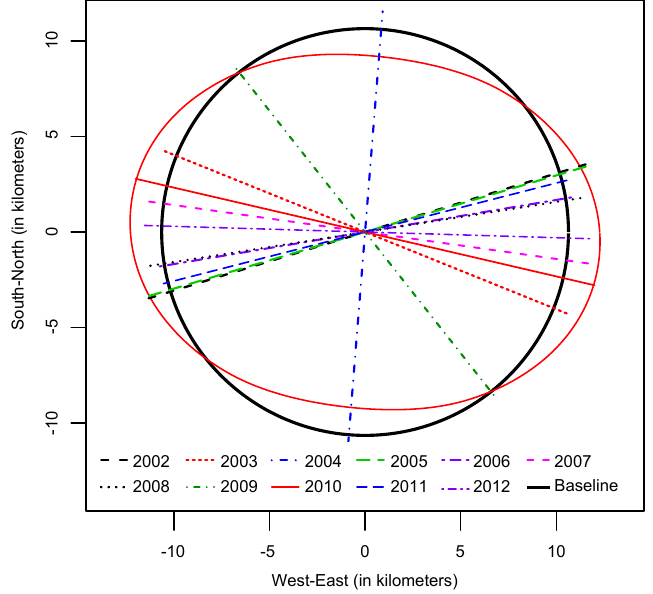}
\caption[]{Baseline contour (thick solid curve), modal axiality contour for the year 2010 (thin solid curve) and diameters of modal axiality contours for the years 2002-2012.}	
 \label{yearwise_fourier_term2}
\end{figure}

In Figure~\ref{yearwise_fourier_term2}, we have plotted the modal axiality contour $g(\hat C^*_2\cos(2(\theta- \hat \phi^*_2)))$ vs. $\theta$ for $\theta\in [0,2\pi]$ in polar coordinates for the year 2010 along with its diameter (solid curve and solid line) and only the diameters of the modal axiality contours for each of years 2002-2009, 2011 and 2012 in different line styles and colors. For reference, we have also included the baseline contour $g(0)$ vs. $\theta$ (thick solid curve). It may be observed that modal axiality is generally found to be around the East-West line. It is interesting to note that predominance of elongation of convective systems in the East-West axis coincides with that of the Inter Tropical Convergence Zone (ITCZ) of the summer monsoon over India \citep{Djuric_1994}.

We have also analyzed the modal directionality of $g(\hat C^*_1\cos(\theta- \hat \phi^*_1))$ corresponding to the first sinusoid (period 2$\pi$) of \eqref{fourier1} in a similar manner. However, no strong mode or modal direction has emerged. We omit the details for brevity.

\section{Comparison of Accuracy of Representation}\label{reconstruction}

We now compare the overall errors of the star-hull based parametric and nonparametric approximations \eqref{fourier3} and \eqref{g3} with that of the corresponding SRVF approximation. The SRVF representation begins with the description of each contour by a pair of continuous functions $x(t)$ and $y(t)$, where parameter $t$ takes value over the periphery of the unit circle. The velocity functions $\dot{x}(t)$ and $\dot{y}(t)$, i.e., the derivatives with respect to $t$, are divided by $\|\dot{x}(t)+\dot{y}(t)\|^{1/2}$ to arrive at the SRVF of that contour, denoted by the vector function $q(t)$. A reparametrization or warping function $\gamma:[0,1]\rightarrow[0,1]$ produces the modified SRVF $q(\gamma(t))\dot{\gamma}^{1/2}(t)$ of the same contour. The distance between two contours is defined as the $L_2$ distance of their SRVFs minimized over $\gamma$, while the Karcher mean of a set of SRVFs is the SRVF with minimum sum of squared distances from the given set. Each of the summand distances occur at some specific $\gamma$, which is used for computation of the Karcher covariance. We have used the R codes used in `fdasrvf' package available from https://github.com/jdtuck/fdasrvf$\_$R for computing the Karcher mean and optimally warped SRVFs of closed curves. We have used the default parameter settings of the package, but bypassed the setting for size invariance. For covariance computations of the SRVF method involving incomplete contours, we have used the technique used in Section \ref{sec:censored} for the star-hull based nonparametric method. 

We use 80\%\ of the complete contours (8921 objects), selected randomly, as training data for all the methods, and the remaining 20\%\ (2219 objects) as test data. The training data are used to obtain the functional principal components for the SRVF and nonparametric star-hull based methods, and to obtain the least squares estimates of the model parameters of the star-hull based parametric method. The overall percentage error of each of the representations is measured by the area of symmetric difference between the actual region under rainfall and the corresponding approximated region (i.e., the area that lies on either one of the regions but not in both), expressed as percentage of the actual area under rainfall, computed from sample contours in the test data set. 

For comparability of the orders of approximation, we use explained percentage of total variance as the benchmark, and opt for two thresholds for this percentage: 90\%\ and 95\%. The number of principal components (or number of cosine terms in the case of the parametric method) needed to achieve this threshold is reported in Table~\ref{table1}, along with different summary statistics of the percentage error of approximation. It may be observed that the star-hull based methods (both nonparametric and parametric) require a much smaller number of basis functions compared to the SRVF method to achieve comparable levels of explained total variation. Further, the three quartiles, the mean and its nominal standard error (computed as the ratio of standard deviation and square root of the sample size) for the two star-hull based methods are comparable to one another and considerably smaller than the corresponding quantities for the SRVF method.
\begin{table}
\vspace{0.4cm}
\caption{Summaries of the error percentages of different representations.}\label{table1}
\vspace{0.5cm}
\centering
\begin{scriptsize}
\begin{tabular}{cccc@{\hskip20pt}cccccc}\hline
$\%$&\multicolumn{3}{c}{SRVF} &\multicolumn{3}{c}{Non-parametric}&\multicolumn{3}{c}{Parametric}\\
variation& $\#$ PCs & Quartiles& Mean (s.e.) & $\#$ PCs & Quartiles & Mean (s.e.)&$d$ &Quartiles & Mean (s.e.)\\\hline
&&Q1=24.5 & & &Q1=18.2& &&Q1=17.9&\\
90 &175 &Q2=31.6& 34.2 (0.32)&7& Q2=22.9& 24.5 (0.18) &4&Q2=22.6&24.0 (0.18)\\
& &Q3=40.6&  &&	Q3=28.8&& &Q3=28.1&\\
\hline
&&Q1=18.3 &  &&Q1=14.1 & & &Q1=13.7&\\
95 &290 &Q2=24.5&27.2 (0.33)&12& Q2=17.6& 19.2 (0.17)&6 & Q2=17.4& 18.9 (0.16)\\
 & &Q3=32.6&  &	&Q3=22.4&& &Q3=22.1&\\\hline
	\end{tabular}
\end{scriptsize}
\end{table}		

It may appear counter-intuitive that an attempt to represent and approximate contour (which the star-hull is) could be more successful than an attempt to represent the exact contour through the SRVF representation. This paradox may be resolved if we consider the large number of basis functions that are needed for the latter method. Each of these basis functions requires computation (with associated computational error) of the corresponding coefficient appropriate for a given contour. The combined errors of the large number of functions may have contributed to the poorer performance of the SRVF method in the test data.

\begin{figure}[]
\centering
\includegraphics[width=6 in] {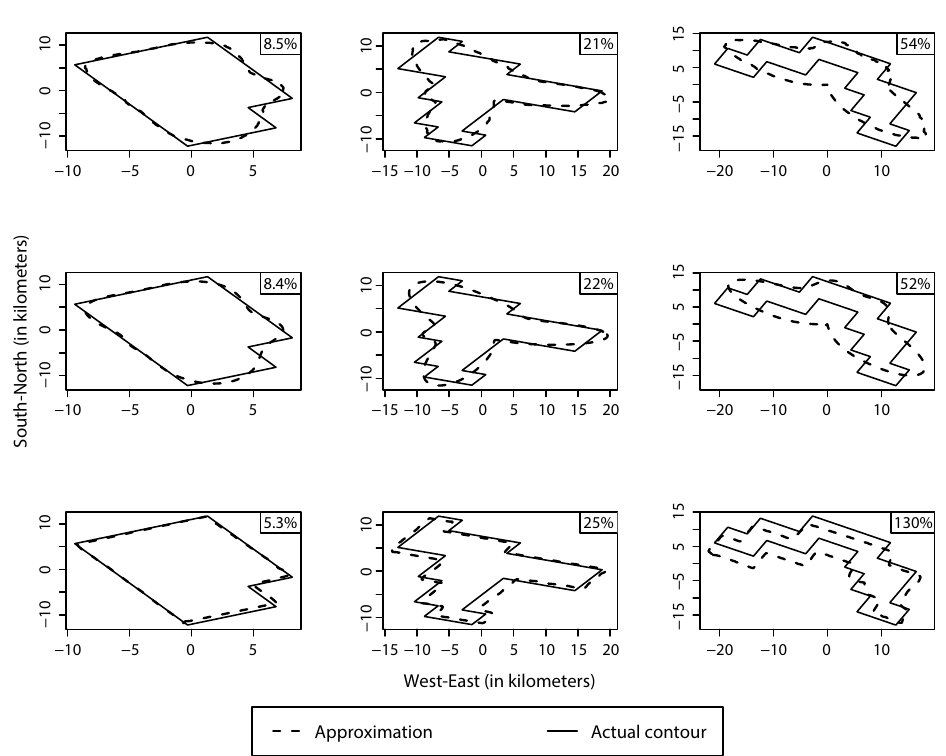}
\caption[]{Examples of contours (solid line) and their reconstructed versions (dashed line) by nonparametric star-hull based method (top), parametric star-hull based method (middle) and SRVF method (bottom). The plots from left to right represent the best, the median and the worst case of approximation error by the SRVF method. Percentage of approximation error with respect to actual area is shown in the top right corner of each panel.}\label{fig13}
\end{figure}

In Figure~\ref{fig13}, we plot some illustrative reconstructions by the three methods, with the number of basis functions chosen as in the bottom part of Table~\ref{table1}. The examples have been selected to show the best (left), the median (middle) and the worst (right) cases of reconstruction in terms of the approximation error. The reconstructions through the two star-hull based techniques are comparable. Notably, the reconstruction through the SRVF method does have some sharp bends, like the original contours. However, the corners of the reconstruction do not necessarily occur near the corners of the original contour. Thus, the proposed methods pick up the turns of the contour in less detail than SRVF, but capture the broad shape better.

\section{Concluding Remarks}

The precipitation areas observed in the TRMM satellite data have diverse and irregular shapes. This fact poses a challenge to modeling, especially for large data sets. In this paper, we have presented an analysis of shapes of contiguous regions under rainfall, through the star-hull envelope of these regions. This approximation, followed by a transformation to correct asymmetry, maps the shape objects to a vector space of functions over $[0,2\pi]$, and makes them amenable to analysis by standard statistical methods.

FPCA of the transformed radial functions of the star-hulls, after adjustment for incomplete contours, revealed certain dominant modes of variation. These modes are easily interpretable and remarkably stable across calendar years. The corresponding basis functions happen to be very close to sinusoids with integer frequencies, which justifies the use of Fourier representation for the star-hulls. This  simple representation has been made possible by the star-hull formulation, which emphasizes the directional aspect of shapes.

It turns out that most of the shape information is captured by a handful of sinusoids. This finding indicates a good potential for data compression. There are also substantial implications to understanding of the MCS precipitation areas, as each sinusoid makes interpretable contributions to the shape.

With the advent of a method to characterize raining systems based on their shape, size and orientation, it might become possible to evaluate the capabilities of Numerical Weather Prediction (NWP) models in simulating such features and to better understand the error characteristics of these models. NWP models are often verified with satellite data such as TRMM data  \citep{Chakraborty}, by using different methods  \citep{Micheas_2007, Ebert_2009}. The phase and  amplitudes of the parametric representation of the contours can be used as alternative features of contiguous precipitation areas, which the verification methods can utilize. Further, the proposed representation might pave the way for studies on how different covariates affect the size, the shape and the orientation of precipitation areas.

Terrestrial radar based measurements are often relied upon for tracking specific raining systems. The proposed representation would permit one to collate such data with satellite based measurements, through temporal evolution of the Fourier coefficients, for improved modeling.

While we used 200 square kilometers as the lower threshold for contiguous areas of positive precipitation, we found the modes of variations to remain stable when the threshold is anywhere between 50 and 1000 square kilometers. Likewise, the pattern did not change much when the threshold for minimum precipitation rate was raised from 0 to 5 mm per hour. These findings indicate that, though our analysis is limited to a certain range of areas under positive precipitation, our method is possibly scalable. Similar analyses of data sets with other resolution and coverage can expand that range, and possibly lead to a more complete understanding of MCS and other convective systems.


\bibliographystyle{apalike}
\bibliography{draft}
\end{document}